\documentclass[twocolumn, showpacs, prl]{revtex4}

\usepackage{graphicx}
\usepackage{dcolumn}
\usepackage{bm}

\begin{document}

\title{Anomalous giant piezoresistance in AlAs 2D electrons with anti-dot lattices}
\date{\today}
\author{O. Gunawan}
\author{T. Gokmen}
\author{Y. P. Shkolnikov}
\author{E. P. De Poortere}
\author{M. Shayegan}
\affiliation{Department of Electrical Engineering, Princeton University, Princeton, NJ 08544}

\begin{abstract}
An AlAs two-dimensional electron system patterned with an anti-dot lattice exhibits a giant
piezoresistance (GPR) effect, with a sign opposite to the piezoresistance observed in the
unpatterned region. We trace the origin of this anomalous GPR to the non-uniform strain in the
anti-dot lattice and the exclusion of electrons occupying the two conduction band valleys from
different regions of the sample. This is analogous to the well-known giant magnetoresistance (GMR)
effect, with valley playing the role of spin and strain the role of magnetic field.
\end{abstract}


\maketitle


Currently there is considerable interest in electronic devices whose operating principles go
beyond the conventional, charge-based electronics. A prime example is the giant magnetoresistance
(GMR) device \cite{BaibichPRL88}, one of the first members of a new class of "spintronic" devices
\cite{PrinzSci98, WolfSci01} whose operation rests on the manipulation of electron's spin degree
of freedom. In certain solids the electrons can reside in multiple conduction band minima (or
valleys) and therefore have yet another, $valley$, degree of freedom. Here we report a giant, low
temperature piezoresistance (GPR) effect in a two-valley AlAs two-dimensional electron system
(2DES) patterned with anti-dot lattices. The effect is among the strongest seen in any system and
allows the detection of minute strains and displacements via a simple resistance measurement. And
yet it is anomalous as it has the opposite sign compared to the conventional piezoresistance found
in multi-valley semiconductors \cite{SmithPR54, ShkolnikovAPL04}. Using magnetoresistance
measurements and numerical simulations, we propose a model that qualitatively explains the
observed GPR effect based on the non-uniform strain and the exclusion of electrons occupying the
two conduction band valleys from different regions of the sample. This is analogous to the
operating principle of the GMR effect: here valley plays the role of spin and strain the role of
magnetic field. These results highlight the fundamental analogy between the spin and valley
degrees of freedom \cite{GunawanPRL06, GunawanNP07} and point to new opportunities in developing
novel "valleytronic" devices whose functionality relies on the control and manipulation of the
electron's valley degree of freedom \cite{GunawanPRB06, RycerzNP06}.

\begin{figure}
\includegraphics[width=\linewidth]{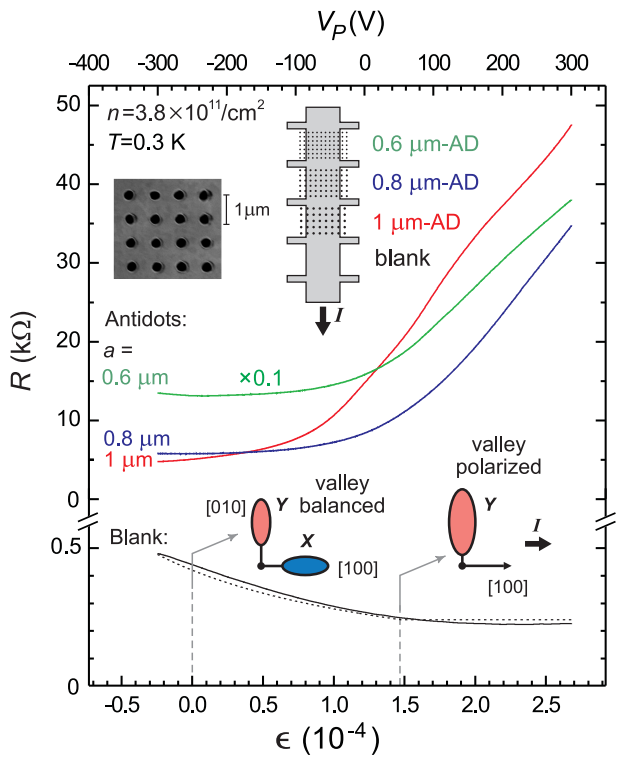} \caption{
(Color online) The piezoresistance of an AlAs 2DES in the un-patterned (blank) region (lower
trace) and in three anti-dot (AD) regions (upper three traces). The dotted line is the best fit to
the piezoresistance in the blank region based on a conventional model. Upper insets: A micrograph
of an AD lattice ($a$=0.8 $\mu$m) and sections of the Hall bar. Lower insets: The orientation and
occupation of the valleys are schematically shown for the blank region at $\epsilon=0$ where the
two valleys are equally occupied and for $\epsilon > 1.5\times10^{-4}$ where all the electrons are
transferred to the $Y$ valley.} \label{FigADPR}
\end{figure}

We performed experiments on a 2DES in a modulation doped, 11 nm-wide AlAs quantum well. In this
system the electrons occupy two in-plane, anisotropic conduction band valleys with elliptical
Fermi contours \cite{PoortereAPL02}, characterized by a heavy longitudinal mass $m_l=1.1 m_0$ and
light transverse mass $m_t =0.2 m_0$, where $m_0$ is electron mass in vacuum. We label these as
$X$ and $Y$ valleys, according to the direction of their major axes, [100] and [010], as shown in
the lower inset of Fig.~\ref{FigADPR} \cite{ShayeganPSSB06}. We patterned a Hall bar along the
[100] direction using standard photo lithography technique. Then, via electron beam lithography
and dry etching using an electron cyclotron resonance etcher we defined three anti-dot (AD)
lattices with periods $a$ = 1, 0.8 and 0.6 $\mu$m in three regions of the Hall bar, and left a
fourth region un-patterned (blank) [see the upper insets of Fig.~\ref{FigADPR}]. Each AD lattice
is an array of holes (ADs) etched to a depth of $\simeq$300 nm into the sample thus depleting the
2DES in the hole area (the 2DES is at a depth of $\simeq$ 100 nm from the top surface). The ratio
$d/a$ for each AD cell is $\sim$1:3, where $d$ is the AD diameter. We also deposited Ti/Au back-
and front-gates to control the total 2DES density ($n$) in the sample. To apply tunable strain, we
glued the sample to one side of a piezo-actuator \cite{ShayeganAPL03}, and monitored the applied
strain using a metal-foil strain-gauge glued to the piezo's other side. We define strain as
$\epsilon =\epsilon_{[100]}-\epsilon_{[010]}$ where $\epsilon_{[100]}$ and $\epsilon_{[010]}$ are
the fractional length changes of the sample along the [100] and [010] directions, respectively.
Note that for $\epsilon > 0$ electrons are transferred from the $X$ valley to the $Y$ valley while
$n$ stays constant \cite{ShayeganPSSB06}. Further fabrication details and characteristics of the
blank region of the particular sample used in this study were reported in
Ref.~\cite{GunawanPRL06}. In particular, at a piezo bias ($V_P$) of -250 V, the $X$ and $Y$
valleys in the blank region are equally occupied (balanced) and, at $n=3.8\times10^{11}$ /cm$^2$
where the data of Fig.~\ref{FigADPR} were taken, electrons are all transferred to  the $Y$ valley
for $V_P > 50$ V ($\epsilon > 1.5\times10^{-4}$); see the lower insets in Fig.~\ref{FigADPR}. The
measurements were performed in a $^3$He cryostat with a base temperature of 0.3 K.

\begin{figure}
\includegraphics[width=65mm]{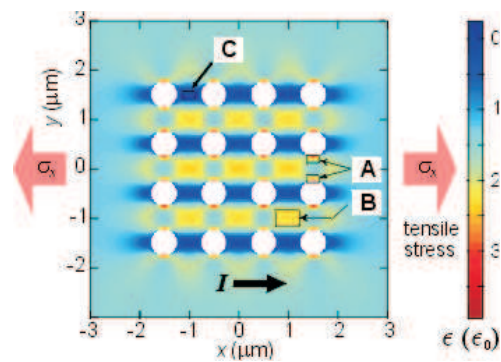} \caption{
(Color online) Finite element simulation of the strain distribution in a 2D medium perforated with
an AD lattice. Boxes A and B highlight the areas of enhanced strain, while box C highlights a
region of reduced strain.} \label{FigADFEM}
\end{figure}

The lower trace in Fig.~\ref{FigADPR} shows the piezoresistance (PR) in the blank region. The PR
exhibits the anticipated behavior: the resistance drops with increasing strain as the electrons
are transferred to the $Y$ valley whose mobility is higher (because of its smaller effective mass,
$m_t$) along the current direction. Beyond the valley depopulation point ( $\epsilon >
1.5\times10^{-4}$), the resistance starts to saturate at a low value as the intervalley electron
transfer ceases. This is the conventional PR effect in AlAs 2DES as has been reported in
Ref.~\cite{ShkolnikovAPL04}. The dotted line in Fig.~\ref{FigADPR} represents the best fit of the
data to a simple model \cite{ShkolnikovAPL04}, which assumes that the valley populations change
linearly with strain, and adds the conductivities of the two valleys to find the total
conductivity; the model also assumes an isotropic scattering time for both valleys and ignores the
inter-valley scattering.

The upper three traces in Fig.~\ref{FigADPR} represent the PR of the AD regions and demonstrate
our main finding. These traces exhibit an $increasing$ resistance as a function of strain,
opposite to the PR in the blank region. The strength of the PR effect is also quite prominent in
the AD regions: indeed, in the 1 $\mu$m-AD region the resistance changes by about ten times for
the range of applied strain while, in contrast, the change for the blank region is only about a
factor of two. Furthermore, for all three AD regions, the PR persists beyond the valley
depopulation point of the blank region ($\epsilon >$ 1.5$\times 10^{-4}$) where the blank region's
PR nearly saturates.

These observations highlight the remarkable difference between the PR effect in the blank and the
AD regions and present an interesting puzzle. As we will now show, it is the presence of the AD
lattice, which significantly modifies the strain distribution in the AlAs 2DES, that leads to the
anomalous PR. To understand the strain distribution in the AD regions we performed a simple
finite-element-method simulation (using FEMLAB) for a plane-strain problem of a 2D medium
perforated with an array of holes. We apply a small tensile stress $\sigma_x$ to the left and
right sides, producing a small amount of strain $\epsilon_0$ at $(x,y)\rightarrow\pm\infty$; in
other words, if there were no AD lattice, the strain would be uniform everywhere with a magnitude
equal to $\epsilon_0$. The result of this simulation is shown in Fig.~\ref{FigADFEM}. There is
clearly a non-uniform strain distribution due to the presence of the AD lattice. In particular,
there are localized regions of enhanced strain (boxes A and B in Fig.~\ref{FigADFEM}), and of
essentially zero strain (box C). For example, in the upper and lower periphery of the AD (box A)
the strain is enhanced by as much as 3$\epsilon_0$. This enhancement by 3$\epsilon_0$ is indeed
indicated by an analytical solution of a 2D plane strain problem with a single hole
\cite{SeeRoylance}. We add that our simulation of Fig.~\ref{FigADFEM} is for a 2D system, however,
we expect that in a system which contains an AD lattice at its top surface, the strain profile is
qualitatively similar to what is shown in Fig.~\ref{FigADFEM}.

But how does a non-uniform strain distribution lead to an increase in resistance? Note that, in
our experiment, positive (negative) strain leads to a valley splitting that favors the $Y$ valley
($X$ valley) occupation. This means that electrons occupying either the $X$ or $Y$ valley feel an
extra, modulated, and confining potential (besides being excluded from the AD hole regions) as
they move through the AD lattice. We believe that it is this potential that profoundly affects the
quasi-ballistic motion of electrons in the AD region and leads to the observed PR. For example,
consider the localized enhancement of positive strain in box A of Fig.~\ref{FigADFEM}. Such strain
depletes the $X$-valley electrons in box A and effectively narrows the width of the channels
(between the holes) through which they have to travel to carry current to the right. In the
remainder of the paper we present evidence from additional measurements and numerical simulations
that lend further support to this picture.


\begin{figure*}
\includegraphics[width=\linewidth]{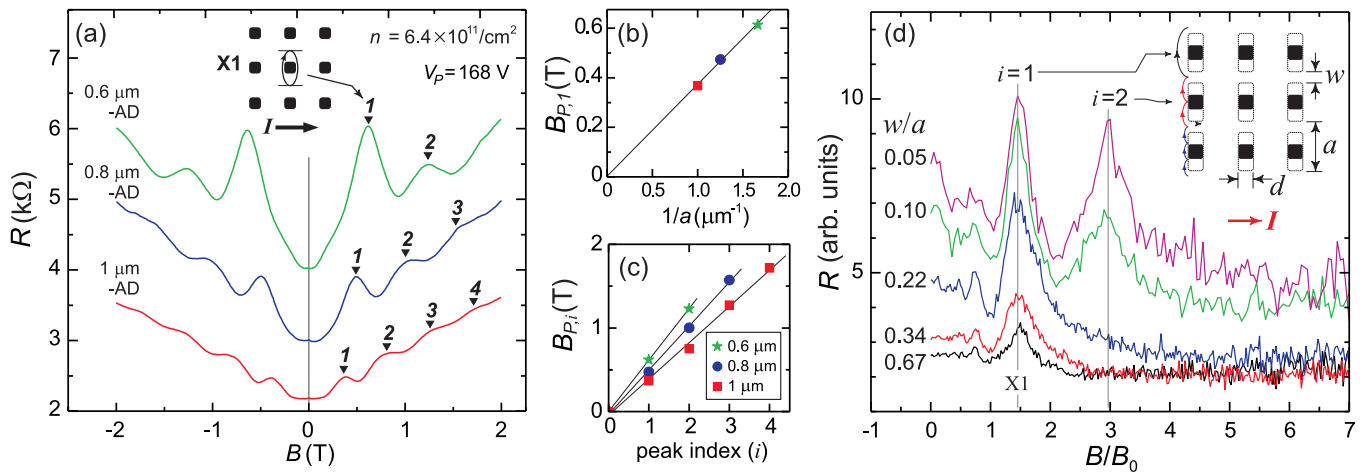} \caption{
(Color online) (a) Magnetoresistance of the AD regions showing the
commensurability peaks labeled with indices $i$ = 1, 2, ... . Inset:
X1 orbit associated with the fundamental ($i$ = 1) peak. (b) The $i$
= 1 peak position $B_{P,1}$ vs. reciprocal AD lattice spacing $1/a$.
(c) Peak positions $B_{P,i}$ vs. index $i$. Each set of data points
fits well to a straight line that goes through the origin. (d)
Numerical simulations of the $X$-valley electron transport through
ADs with variable channel width. $B_0$ is the magnetic field of the
first commensurate orbit (X1) if the Fermi contour were circular.
Inset: Schematic diagram of the "elongated AD" lattice with "channel
width" $w$ and period $a$, and the bouncing orbits that give rise to
sub-harmonic peaks. The black areas are the original AD (holes), the
lines mark the boundaries of the hypothesized, local strain-induced,
"extended AD". } \label{FigMRAll}
\end{figure*}

In Fig.~\ref{FigMRAll}(a) we show resistance vs. perpendicular magnetic field ($B$) traces for the
three AD regions, taken at $V_P = 168$ V ($\epsilon = 2\times10^{-4}$). Each trace exhibits a
series of peaks that we label with indices $i$ = 1, 2, 3, ... . These peaks are associated with
the geometric resonance, or the commensurability, of the cyclotron orbits and the AD lattice
period. Such commensurate orbits are well-known to occur in 2DESs with AD lattices
\cite{WeissPRL91, GunawanPRB07}, although, as we discuss below, the presence of the $i
>$ 1 peaks is unusual. If we plot the positions of the peaks ($B_{P,i}$) vs. their indices as
shown in Fig.~\ref{FigMRAll}(c), we observe that $B_{P,i}$ scales linearly with $i$ for all AD
regions. We refer to the $i = 1$ peak as the fundamental peak and $i = 2, 3, ...$ peaks as the
subharmonic peaks. First we focus our analysis of the fundamental peak and then present a detailed
analysis of the subharmonic peaks.

Figure~\ref{FigMRAll}(b) presents a plot of the fundamental peak
positions $B_{P,1}$ vs. the reciprocal AD lattice spacing $1/a$. The
observed linear dependence is consistent with the geometric scaling
of the peak positions as expected from the relationship: $B_P=\hbar
k_F /e a$, where $k_F$ is the Fermi wave vector of the commensurate
orbit in the direction perpendicular to the current flow
\cite{GunawanPRB07}. In fact, in Ref.~\cite{GunawanPRB07},
systematic measurements and analysis of $B_{P,1}$ as a function of
$n$ and $a$ were made in an AlAs 2DES with AD lattices but without
any applied strain. Both experimental data and simulations showed
that, while the two valleys $X$ and $Y$ could in principle give rise
to two sets of commensurate orbits, it is the X1 orbit
[Fig.~\ref{FigMRAll}(a) inset] that gives rise to the fundamental
magnetoresistance peak. The data in Fig.~\ref{FigMRAll} are
consistent with this finding: If we assign the fundamental peak to
the Y1 orbit (similar to X1 except rotated by 90$^\circ$), we find
that the corresponding $Y$-valley densities deduced from the
positions of these peaks are unphysically large (greater than the
total electron density, $n$, determined from the Shubnikov de-Haas
oscillations). We have repeated such analysis at various $n$ and
$\epsilon$, and have reached a similar conclusion. Therefore, we
surmise that the fundamental peak $B_{P,1}$ is associated with the
X1 orbit.

Now we focus on the sub-harmonic magnetoresistance peaks ($i >$ 1)
observed in Fig.~\ref{FigMRAll}(a). Such peaks are $not$ observed in
the absence of strain, e.g., in the experiments of
Ref.~\cite{GunawanPRB07}, and their presence in
Fig.~\ref{FigMRAll}(a) traces in fact provides clues for the shape
of the potential seen by the $X$-valley electrons in the present
sample. While subharmonic peaks are seldom seen in AD lattices, they
are readily observed in transverse magnetic focusing (TMF)
experiments \cite{vanHoutenPRB89, NiheyAPL90} where ballistic
electrons are emitted through a narrow opening and are collected at
a second narrow opening which is at a relatively large distance
away. Under such conditions, the injected ballistic electrons can
bounce off the TMF barrier one or more times as they follow their
cyclotron orbit trajectories, and magnetoresistance peaks are
observed whenever the emitter-collector distance equals a multiple
integer of the orbit diameter (an illustration of this is shown in
Fig.~\ref{FigMRAll}(d) inset for our structure.) We emphasize that
in TMF structures, the distance between the emitter and collector is
typically larger than the width of the emitter and collector
openings thus allowing bouncing trajectories to occur. Furthermore,
the narrow openings also produce better focusing and therefore sharp
subharmonic peaks.

We hypothesize that the subharmonic peaks observed in
Fig.~\ref{FigMRAll} data arise from an effective narrowing of the
"emitter and collector openings" and an elongation of the effective
AD boundary for the $X$-valley electrons upon the application of
strain as shown in Fig.~\ref{FigMRAll}(d) inset. This is clearly
suggested by the simulations of Fig.~\ref{FigADFEM} where the
$X$-valley electrons are excluded near the lower and upper
boundaries of the ADs (box A) because of the larger local strain.
Several features of Fig.~\ref{FigMRAll} data support this
hypothesis: (1) As shown in Figs.~\ref{FigMRAll}(c) and (d) all the
peak positions are consistent with the orbit diameters being
proportional to $a/i$. (2) The sub-harmonic peaks are pronounced
only when strain is applied: Their amplitudes are indeed small near
zero strain, increase with $\epsilon$, and then saturate. (3) The AD
region with the largest period (i.e. the 1 $\mu$m-AD) has the most
sub-harmonic peaks; this is consistent with the longer AD boundary
allowing more bounces in the electron trajectories.

To further test our conjecture, we performed numerical simulations similar to those used in
Ref.~\cite{GunawanPRB07} but with a variable channel width $w$ to simulate the strain-induced
channel-pinching effect \cite{circle_vs_square}. The results, presented in Fig.~\ref{FigMRAll}(d),
verify our conjecture that a smaller $w$ indeed gives rise to a second sub-harmonic peak that
grows in amplitude relative to the fundamental peak. Note also that, as might be expected,
Fig.~\ref{FigMRAll}(d) simulations show an overall increase in resistance at $B$ = 0 for narrower
channels, consistent with the PR data of Fig.~\ref{FigADPR}.

\begin{figure}
\includegraphics[width=70mm]{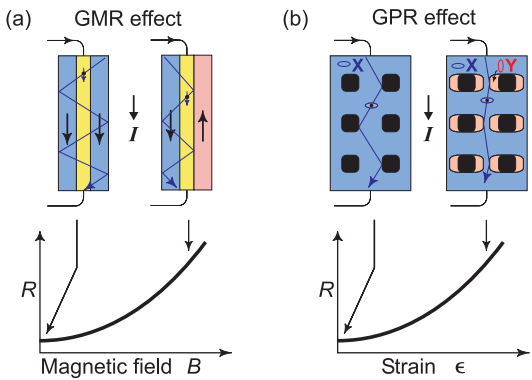} \caption{
(Color online) Comparison between: (a) the GMR effect in a layered magnetic metal sandwich
structure and (b) the GPR effect in AlAs 2DES with an AD lattice.} \label{FigGMRGPR}
\end{figure}

Our experimental data and the numerical simulations combined
strongly suggest a "channel-pinching effect" in the strained AD
lattice for the $X$-valley electrons. Such an effect explains the
emergence of the sub-harmonic commensurability peaks in the
magnetoresistance and also the zero-field PR. However, our model
leaves an important question unanswered: What is the role of the
$Y$-valley electrons which in fact become the majority carriers and
dominate the transport in the blank region with increasing
$\epsilon$ (lower trace in Fig.~\ref{FigADPR})? It is likely that
the non-uniform strain in the AD region creates a strong modulating
potential, limiting the conductivity of the $Y$-valley electrons. An
increase of resistance with the increase of potential modulation
amplitude is indeed common in commensurability oscillation
experiments on surface grating devices \cite{BetonPRB90}. It is
conceivable that although $Y$-valley electrons are favored at high
strains, their contribution to the overall AD lattice conductivity
is suppressed by the strongly modulated potential.

The above channel-pinching picture suggests a remarkable resemblance between our GPR effect and
the GMR effect, observed in thin-film structures composed of alternating layers of ferromagnetic
and non-magnetic materials \cite{PrinzSci98, EgelhoffJVST99}. A comparison of these two effects is
schematically illustrated in Fig.~\ref{FigGMRGPR}. In each structure, the reversal of polarization
of the magnetization (spin) or valley in the regions adjacent to the active channel due to either
external magnetic field ($B$) or applied strain ($\epsilon$) leads to a narrower effective channel
width (extra confinement) and possibly additional scattering (for both $X$ and $Y$-valley
electrons) both of which lead to higher resistance.

Regardless of its origin, the GPR exhibited by our AD lattices
reveals the extreme sensitivity of their resistance to strain.  The
data of the 1 $\mu$m AD lattice, e.g., yield a maximum strain gauge
factor, $\kappa$, defined as the fractional change in sample
resistance divided by the fractional change in sample length, of
over 20,000.  This is by far larger than $\kappa \simeq 2$ of
standard, metal foil gauges, and is among the largest $\kappa$
reported for any solid state material.  Our structure may find use
as an extremely sensitive, low-temperature PR strain sensor to
detect ultrasmall forces and distances.  Using a simple resistance
measurement, we were able to detect strains down to $2\times10^{-8}$
with our samples \cite{GunawanThesis07}. Given that the spacing
between our Hall bar resistance contacts is 40 $\mu$m, this strain
translates to a displacement of 8$\times$10$^{-4}$ nm (about 1/50 of
the Bohr radius)! This sensitivity could be further improved by
designing AD lattices with optimized shapes and sizes, and using
more sophisticated techniques to measure resistance changes.

We thank the ARO and NSF for support, and K.\ Vakili for illuminating discussions.


\begin{references}
\bibitem{circle_vs_square}
For simplicity and also because the drawn shape of each AD was
intended to be a square, in the simulations of
Fig.~\ref{FigMRAll}(d) we used square and rectangular shapes for the
ADs as indicated in the figure. We repeated the simulations with
circular and elliptical AD shapes and obtained qualitatively similar
results.
\end{references}

\end{document}